\documentclass[preprint]{emulateapj}

\def\la{\lambda}

\usepackage[dvips]{color}

\usepackage{aas_macros} 
\usepackage{here}
\usepackage{amsmath,amssymb}
\slugcomment{}
\shortauthors{Hirano et al.}
\shorttitle{Measurements of Stellar Inclinations for KOI Systems}
\begin{document}
\title{Measurements of Stellar Inclinations for Kepler Planet Candidates}
\author{
Teruyuki Hirano\altaffilmark{1}, 
Roberto Sanchis-Ojeda\altaffilmark{2},
Yoichi Takeda\altaffilmark{3},
Norio Narita\altaffilmark{3}, 
Joshua N.\ Winn\altaffilmark{2},
Atsushi Taruya\altaffilmark{1,4,5}, and
Yasushi Suto\altaffilmark{1,4,6}
} 
\altaffiltext{1}{Department of Physics, The University of Tokyo, 
Tokyo 113-0033, Japan}
\altaffiltext{2}{Department of Physics, and Kavli Institute 
for Astrophysics and Space Research, Massachusetts Institute of Technology,
Cambridge, MA 02139}
\altaffiltext{3}{National Astronomical Observatory of Japan, 
2-21-1 Osawa, Mitaka, Tokyo, 181-8588, Japan}
\altaffiltext{4}{Research Center for the Early Universe, School of Science, 
The University of Tokyo, Tokyo 113-0033, Japan}
\altaffiltext{5}
{Institute for the Physics and Mathematics of the Universe (IPMU), 
The University of Tokyo, Chiba 277-8582, Japan}
\altaffiltext{6}{Department of Astrophysical Sciences, 
Princeton University, Princeton, NJ 08544}

\email{hirano@utap.phys.s.u-tokyo.ac.jp}
\begin{abstract}
We present 
an investigation of spin-orbit angles
for planetary system candidates reported by Kepler. By combining the rotational period
$P_s$ inferred from the flux variation due to starspots and the
projected rotational velocity $V\sin I_s$ 
and stellar radius
obtained by a high resolution
spectroscopy, we attempt to estimate the inclination $I_s$ of the 
stellar spin axis with respect to the line-of-sight.  For transiting
planetary systems, in which planetary orbits are edge-on seen from us,
the stellar inclination $I_s$ can be a useful indicator of a spin-orbit
alignment/misalignment.  We newly conducted spectroscopic observations
with Subaru/HDS for 
15
KOI systems, whose lightcurves show periodic flux
variations.  After detailed analyses of their lightcurves and spectra,
it turned out that some of them are binaries, or the flux variations
are too coherent to be caused by starspots, 
and consequently we could constrain stellar
inclinations $I_s$ for eight systems. 
Among them, KOI-262 and 280 are in good agreement with $I_s=90^\circ$
suggesting a spin-orbit alignment, while at least one system, KOI-261, shows 
a possible spin-orbit misalignment. We also obtain a small $I_s$ for KOI-1463, 
but the transiting companion seems to be a star rather than a planet. 
The results for KOI-257, 269, 367, and 974 are ambiguous, and can be explained 
with either misalignments or moderate differential rotation.
Since our method can be applied to any system having starspots regardless 
of the planet size, future observations will 
allow for the expansion of
the parameter space in which the spin-orbit relations are investigated.
\end{abstract}

\section{Introduction\label{s:intro}}\label{s:intro}

The standard formation theory of close-in gas giants (hot-Jupiters)
suggests that they form outside of the so-called ``snow-line'', located
at a few AU away from the host star, and subsequently migrate inward
\citep[e.g.,][]{1996Natur.380..606L, 2009AREPS..37..321C,
2010exop.book..347L}.  While migration processes such as disk-planet
interactions (type I or II migration) predict small orbital
eccentricities of planets and small stellar obliquity (i.e. the angle
between the stellar spin axis and the planetary orbital axis), dynamical
processes including planet-planet scattering and/or Kozai cycles might
produce large values for both eccentricity and stellar obliquity
\citep[e.g.,][]{2003ApJ...589..605W, 2007ApJ...669.1298F,
2008ApJ...678..498N, 2008ApJ...686..580C, 2011ApJ...742...72N}.

Measurements of the Rossiter-McLaughlin effect (hereafter, the RM
effect), which is an apparent radial velocity anomaly during a planetary
transit, provide invaluable information to better understand the
planetary migration process \citep{2005ApJ...622.1118O, 2005ApJ...631.1215W, 
2007PASJ...59..763N, 2007A&A...474..565A, 2008A&A...488..763H, 2010A&A...524A..25T, 
2011ApJ...742...69H}.  Through
the RM effect, one can measure the sky-projected angle $\la$ between the
stellar spin axis and planetary orbital axis, which is of importance to
distinguish among the possible migration channels.  

So far, more than 50 transiting systems have been investigated to
estimate $\la$, and many interesting correlations among spin-orbit
angles and physical properties of planets and host stars have been
proposed
\footnote{The list of RM measurements is available at
http://www.aip.de/Peopler/RHeller}.  
\citet{2010ApJ...718L.145W} pointed out that a substantial
spin-orbit misalignment tends to be observed around hot stars (whose
effective temperatures $T_\mathrm{eff}\gtrsim 6250$ K). This fact might
be related to the tidal evolution of host star's obliquities.
\citet{2011A&A...534L...6T} also found a correlation between the stellar
ages and obliquities, claiming that spin-orbit misalignments are
observed around younger systems with ages less than 2.5 Gyr.  
This trend is consistent with what \citet{2010ApJ...718L.145W} found, suggesting 
that tidal interactions
between the host stars and close-in giant planets, formed after some
dynamical processes such as planet-planet scattering, gradually damp
the obliquity of host stars close to $0^\circ$.

It should be noted that measurements of the RM effect are only feasible
for rather bright stars ($V\lesssim 12$) with giant transiting planets.
While detections of the RM effect for a super-Neptune were
reported \citep{2011PASJ...63S.531H, 2010ApJ...723L.223W}, 
those for smaller planets ($R_p\lesssim 0.5 R_J$) are still
challenging.  Nevertheless, in order to discuss planetary formation
and migrations, it is of great interest to investigate the spin-orbit
relations for systems with Neptune-sized or even Earth-sized exoplanets,
which are reported to be more abundant than jovian planets
\citep{2011ApJ...736...19B, 2011arXiv1109.2497M}.

\begin{figure}[t]
\begin{center}
\includegraphics[width=9cm,clip]{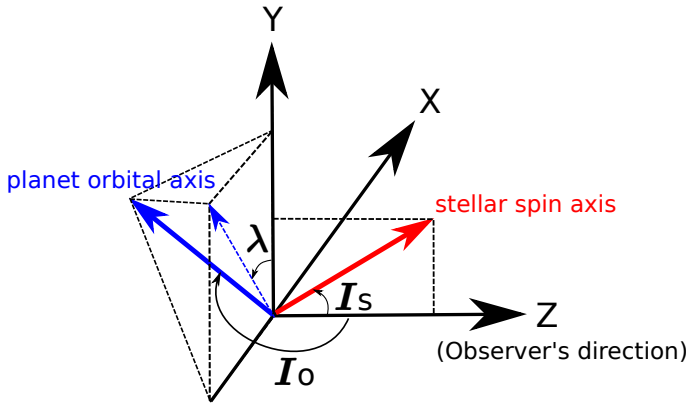} 
\caption{
Schematic figure for the spin and orbital axes. In this figure, the $x-y$ plane 
indicates the sky plane and the $z$ axis points toward us. The planet's orbital
axis is indicated by the blue arrow and its projection onto the sky plane
is shown by the dashed arrow. The red arrow is the stellar spin axis, which
is located in the $y-z$ plane.
}\label{spin-orbit_angles}
\end{center}
\end{figure}

In order to measure spin-orbit relations for such smaller planets, we
focus on the Kepler photometry in this paper. The Kepler mission is an
ambitious and very productive project designed to ``determine the
frequency of Earth-size planets in and near the habitable zone of
solar-type stars''.  As of March 2012, more than two thousand planetary
candidates were announced by the Kepler team \citep{2010Sci...327..977B,
2011ApJ...736...19B, 2012arXiv1202.5852B}, 
and each of those stars
having planetary candidate(s) is called Kepler Object of Interest (KOI). 
Among the published Kepler
light curves, there are many systems that show periodic flux variations
most likely due to starspots on the stellar surface
\citep{2011AJ....141...20B}.  For those systems, a period analysis
enables us to infer the rotational period $P_s$ of the star. That
rotational rate, along with the stellar radius $R_s$, can be directly
translated into the rotational velocity at the equator of the star,
$V_\mathrm{eq}$ 
in the absence of differential rotation. 
If we compare $V_\mathrm{eq}$ with the projected
rotational velocity $V\sin I_s$ estimated by a spectroscopic
observation, we can constrain the stellar inclination $I_s$, which is
defined as the angle between the line-of-sight and the axis of the
stellar rotation (see Figure \ref{spin-orbit_angles}).  Since transiting
planetary systems nearly have edge-on orbits seen from our location, the
orbital inclination $I_o$ (the angle between our line-of-sight and the
planetary orbital axis) should be close to $90^\circ$.  Therefore, a
significant deviation of $I_s$ from $90^\circ$ implies a spin-orbit
misalignment.  This method to constrain the stellar obliquities will be
described in detail in Section \ref{principle}.

The statistics of the spin-orbit angles along the line-of-sight has been
discussed by \citet{2010ApJ...719..602S}.  Using the empirical relation
among the stellar ages, masses, and rotational periods, he statistically
discussed the rotational velocity of the stars hosting transiting planets.
He found that some of the transiting systems have smaller
$V\sin I_s$ than expected for the case of spin-orbit alignment, and
pointed out that it is most likely to be evidence of spin-orbit
misalignments along the line-of-sight.  In contrast to his analysis that
relies on an empirical relation to estimate the rotational period $P_s$
of the planet hosting stars, we attempt to derive it more directly for
each of the KOI systems using the precise Kepler photometry. In order to
constrain $I_s$, we newly conducted high resolution spectroscopic
observations and obtained spectra for 15 KOI systems using the
Subaru telescope.

The remaining sections are organized as follows. In Section
\ref{principle}, we describe the basic method to estimate the spin-orbit
angle along the line-of-sight in detail and discuss the pros and cons of
the present method in comparison with the RM effect. We briefly describe
the spectroscopic observations with Subaru in Section
\ref{s:obs}. Section \ref{s:result} presents the photometric and
spectroscopic analyses and their results.  The correlations between the
stellar inclinations and other system parameters are discussed in
Section \ref{s:discussion}.  Finally, Section \ref{s:conclusion} is
devoted to summary and future prospects.

\section{Principle \label{principle}}\label{principle}

\begin{figure}[t]
\begin{center}
\includegraphics[width=9cm,clip]{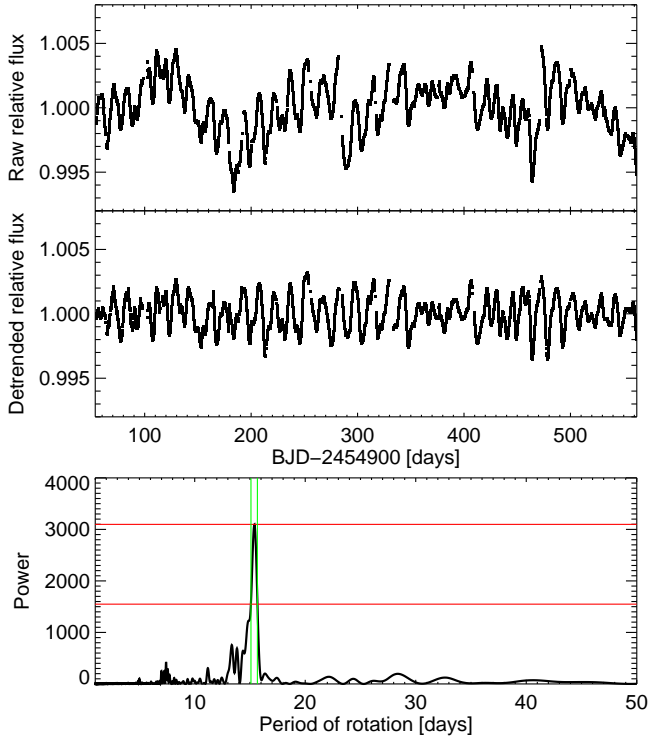} \caption{ {\it
Top.} - Quarter by quarter mean normalized raw out-of-transit flux of
KOI-261 (PDC~SAP$\_$FLUX). Includes quarters 0 through 6, except quarter
5 where this target was not observed. {\it Central.}- Detrended flux of
the star, obtained with the method described in Section \ref{s:result}.
{\it Bottom.} - Lomb-Scargle periodogram of the detrended flux.  The
solid vertical lines represent the points where the power is half of the
maximum power. The final value is taken to be the mean of these points,
and the error to be half of their difference.  }\label{koi261}
\end{center}
\end{figure}

If there exists a spot on the surface of a star, the lightcurve of the
star exhibits a periodic variation due to the stellar rotation. A period
analysis of the lightcurve enables us to estimate the rotational period
$P_s$ of the star.  Figure \ref{koi261} shows an example of the Kepler
lightcurve (KOI-261) and its periodogram. The peak in the periodogram
most likely reflects the rotational period $P_s$, which is estimated as
$P_s=15.4\pm 0.3$ days.  Once $P_s$ is estimated, the stellar
inclination $I_s$ is estimated by the following relation:
\begin{eqnarray}
I_s=\arcsin \left\{\frac{P_s (V\sin I_s)_\mathrm{spec}}{2\pi R_s}\right\},
\label{I_s}
\end{eqnarray}
where $(V \sin I_s)_\mathrm{spec}$ and $R_s$ are the projected
rotational velocity and the radius of the star, respectively.  Both of
these quantities are estimated via spectroscopy under the assumption
that the star is rigidly rotating.  The impact of differential rotation
will be discussed in Section \ref{s:discussion}. 

Since the configuration of the transiting system has an edge-on orbit
seen from us (with the orbital inclination $I_o\gtrsim 85^\circ$), a
small value of $I_s$ implies a possible spin-orbit misalignment in the
system, regardless of the sky-projected spin-orbit angle $\la$ (see
Figure \ref{spin-orbit_angles}).  The 3D angle $\psi$ between the
stellar spin axis and planetary orbital axis is associated with $I_s$,
$I_o$, and $\la$ by the following equation \citep{2009ApJ...696.1230F}:
\begin{eqnarray}
\cos\psi=\sin I_s\cos\la\sin I_o+\cos I_s\cos I_o.
\label{eq:psi}
\end{eqnarray}

The present method has several advantages that we mention below.
Although observations of the RM effect have enabled us to discover the
spin-orbit misalignment for the first time and revealed its possible
patterns in terms of stellar and planetary properties
\citep{2008A&A...488..763H, 2009PASJ...61L..35N, 2009ApJ...703L..99W,
2010ApJ...718L.145W, 2011A&A...534L...6T}, the methodology is feasible
only for sufficiently bright stars ($V\lesssim 12$) with giant
transiting planets ($R_p\gtrsim 0.5 R_J$).  In addition, a measurement
of $\la$ requires spectroscopic observations throughout a complete
transit. Therefore the observation is time-critical and also
time-consuming, in particular for those systems in which the semi-major
axis of the planet is large and the transit duration is long.

In contrast, the current method of measuring spin-orbit relations uses
the effect of stellar spots induced by rotation on the flux of the star,
and thus requires only one spectroscopic observation independently of
the size of planets and their semi-major axes. Therefore the present
method is more efficient in increasing the number of samples, and should
expand the parameter space in which spin-orbit relations are
investigated.

In systems where spots are present, there is a possibility that the
planet passes in front of one of them during a transit.  The recurrence
of these spot-crossing events at consecutive transits proves that the
system has a low spin-orbit angle \citep{2011ApJ...733..127S,
2011ApJ...740L..10N, 2011ApJS..197...14D}.  
However, to locate the spot anomalies for this spot-crossing method requires 
a high signal-to-noise ratio for the transit lightcurve, which is achievable only for large 
planets or very bright stars. Also it would be more suitable for cool stars where 
active regions tend to be larger and produce more visible spot anomalies. 
In addition to this, cool stars generally have a lower $V\sin I_s$, which complicates 
the use of our method.

On the other hand, measuring the spin-orbit relation along the
line-of-sight with the present technique has a few shortcomings.  First,
because of the shape of the sine function, the uncertainty for $I_s$
tends to be larger when $\sin I_s$ is close to unity.  Second, it is
impossible to distinguish the state of $I_s$ from that of $\pi - I_s$,
which gives exactly the same value of $\sin I_s$.  This is a marked
contrast with the RM measurements, through which we can clearly
distinguish between the prograde and retrograde orbits at least on the sky. 

Finally, types of stars to which we can apply our present technique are
fairly limited.  According to the survey by \citet{1982PASP...94..934R},
the flux variations due to dark starspots are only seen in Sun-like
stars whose effective temperatures are 
up to 6400 K, with a
photometric precision of $\sim 3$ mmag. Therefore, if we observe a
periodic flux variation with a large amplitude for an early type star
hotter than 6500 K, 
it is more likely to be spurious, possibly reflecting a
flux contamination by a companion star or faint background sources. 
Therefore the present method is complementary to the RM effect, 
in the sense that RM measurements are applicable
to even hotter stars \citep[e.g.,][]{2010MNRAS.407..507C}.

There are some systems in which spots were used to estimate
rotational periods of the planet-hosting stars 
\citep[e.g.,][for CoRoT-2 and Kepler-17, respectively]{2009A&A...493..193L, 2011ApJS..197...14D}.
Specifically, \citet{2011A&A...533A.130H} applied the present technique to 
investigate the spin-orbit relations to the CoRoT-18 system, and obtained a weak constraint
on the stellar inclination as $I_s=70^\circ\pm 20^\circ$. This was in 
good agreement with the small value of the sky-projected spin-orbit angle 
($\lambda=10^\circ\pm 20^\circ$), implying a 3D spin-orbit alignment. 
In this paper, we systematically apply this technique to some of the
KOI systems showing periodic flux variations. 

\section{Observation \label{s:obs}}\label{s:obs}

We applied the following three criteria to all the KOI systems in the
February 2011 data release \citep{2011ApJ...736...19B}, 
and selected 15 KOI
systems (KOI-4, 42, 100, 257, 258, 261, 262, 269, 279, 280, 302,
367, 974, 1020, 1463); 1) the lightcurve exhibits flux variations
by a visual inspection, 2) the (photometrically estimated) effective
temperature is higher than $\sim 6000$ K, and 3) the Kepler magnitude is
brighter than $K_p\sim 13.0$. The second criterion is adopted since
hotter stars are likely to have larger $V\sin I_s$ and so the relative
statistic and systematic errors in $I_s$ become smaller.

In order to derive spectroscopic parameters for those KOI systems, we
conducted spectroscopic observations with the High Dispersion
Spectrograph (HDS) installed on the Subaru telescope, located in
Hawai'i.  The observations were performed on June 12, July 23, 24, and
October 19, 2011 (UT). We employed the I2a observing mode, which covers
the wavelength regions of $4950~\mathrm{\AA}$ to $6150~\mathrm{\AA}$
(blue CCD) and $6400~\mathrm{\AA}$ to $7550~\mathrm{\AA}$ (red CCD). The
spectral resolutions were set as $R\sim 90,000$ on June 12 and July 24,
$R\sim 60,000$ on July 23, and $R\sim 110,000$ on October 19,
respectively. 
The resolution for each target is summarized in the middle column of Table \ref{table0}.
The seeing on each observing nights was typically $0.5-0.8$ arcseconds.

On each of those observing nights, we obtained the reference spectrum of
the flat lamp transmitted through the Iodine cell, using the same setup
(spectral resolution).  These reference spectra were used to reproduce
the instrumental profile for each spectral resolution, and then play a
crucial role in estimating the projected rotational velocity $V\sin
I_s$ for slowly rotating stars. 

We reduced the raw data of each spectrum using the standard IRAF
procedure.  The resulting spectra have the typical signal-to-noise ratio
(SNR) of $80-120$ per pixel after extracting the 1D spectra.

\section{Analyses and Results \label{s:result}}\label{s:result}
\subsection{Estimation of Rotational Periods}\label{s:periods_estimate}

In order to determine the periods of photometric variations, we used all
public data available from the MAST archive for the 15 targets. In most
cases, quarters 0 through 6 were available, corresponding to 500 days of
observations in total. In this analysis, only the long cadence
observations were used, which proved to be sufficient to study the flux
variability with a timescale of a few days. Since the Pre-search Data Conditioning
(PDC) pipeline is known to remove partially or totally the stellar flux variability on the
timescales we are interested in \citep{2010ApJ...713L.120J}, we decided
to use the raw flux, named SAP\_FLUX in the version 2.0 of the FITS
files delivered by the Kepler team.
Note that the newest improvements on the Kepler pipeline 
\citep{2012arXiv1203.1382S, 2012arXiv1203.1383S} also seem to allow 
for a fast identification of the real astrophysical noise, but we simply
apply the method below in the present paper. 

We used the linear ephemeris of all planet candidates published in the
February 2011 data release \citep{2011ApJ...736...19B} to locate all
transits and remove them from the flux series. Then, using a smoothed
version of the flux where each point is the average of the previous and
next 10 points, we estimate the standard deviation of the flux, and we
applied 3 sigma clipping to the data in order to remove outliers. Raw
Kepler data suffer from several well known instrumental artifacts
\citep{2010ApJ...713L..87J} that would not be corrected by the 3 sigma
clipping.  Some of them only affect a certain interval of the
observations, for example the changes in flux due to temperature drifts
after the telescope points to the Earth. 

There are also long term trends due to the constant movement of the
targets on the CCD. 
In order to remove them, we detrend the data using
the cotrending basis vectors, following the instructions from the Data
Release 12 Notes. These cotrending vectors are constructed to capture
most of the flux variability caused by instrumental artifacts in each CCD.
The mean-subtracted, mean-normalized flux of each target can be
considered as a superposition of the astrophysical flux of the star plus
a linear combination of these orthonormal basis of cotrending
vectors. Assuming that the astrophysical flux is orthogonal to the
instrumental noise, we can estimate the coefficients of the linear
combination by simply taking the vector product of the mean-subtracted,
mean-normalized raw flux with as many cotrending vectors as needed to
clean the lightcurve. After this removal, we added back the previously
subtracted mean, to then normalize by the mean each quarter to avoid
differences in flux from quarter to quarter.

We applied the above procedure to the 15 targets, 
and studied the final flux lightcurve. Visual inspection of the light
curves confirmed the existence of flux variability due to starspots on 13 
out of the 15 cases, with periodic signals 
with amplitudes and phases that evolve with time.
KOI-258 and KOI-302 show a very clear but coherent
variability that does not change shape with time. Such strictly periodic
signals are unlikely to be caused by spots, especially in the case of
KOI-258 where the period of the signal is equal to the orbital period of
the candidate.  This strongly suggests that the periodic dimming of KOI-258 is
likely to be caused by a background binary.

\begin{table}[t]
\caption{
Spectral resolutions adopted in spectroscopic observations
by Subaru/HDS
and rotational periods estimated by the Kepler photometry.
}\label{table0}
\begin{center}
\begin{tabular}{ccc}
\hline
System & $R$ &$P_s$ (days)\\\hline\hline
KOI-4 & $\sim110000$ & $5.65\pm 0.03$\\
KOI-42& $\sim90000$ & $20.84\pm 0.37$ \\
KOI-100 & $\sim90000$ &$1.132\pm0.002$\\
KOI-257&$\sim 90000$ &$7.846\pm0.052$\\
KOI-258& $\sim 90000$& too coherent \\
KOI-261& $\sim 110000$&$15.38\pm0.30$ \\
KOI-262 & $\sim 90000$&$8.171\pm1.218$\\
KOI-269 & $\sim 90000$ &$5.351\pm0.136$\\
KOI-279 & $\sim 90000$ &$20.92\pm0.93$\\
KOI-280 & $\sim 90000$ &$15.78\pm2.12$\\
KOI-302 & $\sim 60000$ & too coherent \\
KOI-367 & $\sim 90000$ &$27.65\pm3.56$\\
KOI-974 & $\sim 110000$ &$10.83\pm0.12$\\
KOI-1020 & $\sim 90000$ & $10.91 \pm 1.06$\\
KOI-1463 & $\sim 60000$ &$6.042\pm0.042$\\
\hline
\end{tabular}
\end{center}
\end{table}

In order to estimate the periods of rotation for the rest of the stars, we
use a lomb-scargle algorithm and analyze the power spectra of the
stellar fluxes. A high peak is expected to happen at the period of
rotation of the star, although a strong peak can also appear at half the
period if several spots are present. 
The peak can also be rather wide if differential rotation is present, or even 
be composed of several peaks. However, visual inspection can also reveal 
important information about the periodic behavior of the flux series, and it 
helped to identify a few cases where indeed the highest peak corresponded 
to half the period of rotation. In these cases, two similar spots, with opposite 
stellar longitudes, induce flux variations with twice the frequency. The evolution 
of the size of both spots, which creates asymmetries between the flux 
variations induced by both, allows us to identify them unambiguously. 

The final values of the period of rotation and their errors are obtained by
studying the proximity of the strongest peak, and summarized in the right column of
Table \ref{table0}. We 
adopt the full-width at half maximum (FWHM) of
the peak as the 1-sigma error, with the center of the interval being the
rotation period. Note that if a differential rotation produces several
peaks with similar amplitudes at similar periods, the resulting merged
peak in the periodogram is broadened, and the above assigned error
becomes large. Thus, the presence of stellar differential rotation makes
the estimate of the period less accurate. 

\subsection{Estimation of Spectroscopic Parameters}

Following \citet{2002PASJ...54..451T, 2005PASJ...57...27T}, we analyze
each spectrum and estimate the effective temperature $T_\mathrm{eff}$,
surface gravity $\log g$, microturbulence $\xi$, and metallicity [Fe/H]
by measuring the equivalent widths of Fe I and Fe II lines.  In order to
accurately estimate $V\sin I_s$ with avoiding any systematic effect, we
numerically integrate each component on the stellar disk Doppler-shifted
due to stellar rotation and macroturbulence.  In doing so, we also
convolve the intrinsic (thermal motion + microturbulence) profile with
the rotational and macroturbulence broadening function and the
instrumental profile to reproduce the observed spectrum. 

For macroturbulence, we adopt the radial-tangential model
\citep{2005oasp.book.....G}.  For the given macroturbulence dispersion
$\zeta_\mathrm{RT}$ and the stellar limb-darkening parameter ($\epsilon
=0.6$), we determine the best-fit solution of $V\sin I_s$ for each of
the spectra. Since the macroturbulence dispersion $\zeta_\mathrm{RT}$ is
not well understood, especially for hotter stars, we try several
different values of $\zeta_\mathrm{RT}$. For the final result, we adopt
the following empirical expression by \citet{2005ApJS..159..141V}:
\begin{eqnarray}
\zeta_\mathrm{RT}
=\left(3.98+\frac{T_\mathrm{eff}-5770~\mathrm{K}}{650~\mathrm{K}}
\right)~\mathrm{km~s}^{-1},
\label{zetaRT}
\end{eqnarray}
and estimate the systematic uncertainty for $V\sin I_s$ by changing
$\zeta_\mathrm{RT}$ by $\pm 15\%$ from Equation (\ref{zetaRT}) for cool
stars ($T_\mathrm{eff}\leq6100$ K) based on the observed distribution of
$\zeta_\mathrm{RT}$ \citep[see Figure 3 in][]{2005ApJS..159..141V}.  For
the case of hotter stars ($T_\mathrm{eff}>6100$ K), however, the
macroturbulence is not intensely investigated and thus we conservatively
estimate the systematic error for $V\sin I_s$ by changing
$\zeta_\mathrm{RT}$ by $\pm 25\%$.  The statistical errors in fitting
each spectrum are generally smaller than the systematic errors arising from
different values of $\zeta_\mathrm{RT}$.

As for the instrumental profile, we basically adopt Gaussian broadening
functions whose FWHMs correspond to each of the spectral resolutions
adopted in the observations ($R\sim 60,000$, 90,000, and 110,000).  In
the case of slowly rotating stars ($V\sin I_s\lesssim 5$ km s$^{-1}$),
however, we found that the shape of the instrumental profile, which is
slightly different from Gaussian, sometimes affects the estimate of
$V\sin I_s$. Thus, we convolve the actual shapes of the instrumental
profile estimated by the reference transmission spectrum of the Iodine
cell for those slow rotators (i.e., KOI-261 and KOI-367).

\citet{1995PASJ...47..337T} applied this procedure to the high-resolution solar 
flux spectrum and obtained the solar spin velocity of $V\sin I_s=2.00\pm 0.34$ km s$^{-1}$.
Since the angular velocity around the solar equator is about 14$^\circ~\mathrm{day}^{-1}$
from the observations of spatially resolved spots \citep{2005SoPh..229...35R}, 
we obtain $V_\mathrm{eq}\approx 2.0$ km s$^{-1}$ for the Sun. 
We also applied this technique to the spectrum of HAT-P-11, whose rotational velocity is reported
to be very small \citep[$V\sin I_s=1.00_{-0.56}^{+0.95}$ km s$^{-1}$,][]{2010ApJ...723L.223W} 
from the measurement of the RM effect, independently of the spectral line analysis. 
Although the resolution of HAT-P-11's spectrum that we analyzed is relatively 
low ($R\sim 45000$), we obtained $V\sin I_s=1.79\pm 0.65$ km s$^{-1}$ from the line analysis 
(including the convolution of the instrumental profile), 
which is consistent with the result estimated by the RM measurement. 
These two tests (for the Sun and HAT-P-11) validate the present technique to estimate
$V\sin I_s$ for slow rotators. 
Of course, they do not guarantee the extrapolation 
of even smaller $V\sin I_s$ ($\lesssim 1.0$ km s$^{-1}$, e.g., KOI-261), but 
for such stars, we can safely rule out a large 
$V\sin I_s$ ($\gtrsim 2$ km s$^{-1}$) with a high confidence level. 

\begin{table*}[t]
\caption{
Spectroscopic Parameters. Starred systems have companion stars, 
most likely causing contaminations in spectra.
}\label{table1}
\begin{tabular}{ccccccccc}
\hline
System &$T_\mathrm{eff}$ (K) & $\log g$ & [Fe/H] 
& age (Gyr) & $M_s$ ($M_\odot$) & $R_s$ ($R_\odot$) 
& $V\sin I_s$  (km s$^{-1}$)
& $V_\mathrm{eq}$ (km s$^{-1}$) \\\hline\hline
KOI-42$^\star$ &$6512 \pm58$ & $4.542\pm 0.090$& $0.114\pm 0.050$
& $<0.06$ & $1.373\pm0.022$& $1.293_{-0.044}^{+0.045}$
& $13.40\pm0.22$&$3.14\pm0.12$ \\
KOI-257&$6218\pm28$&$4.286\pm 0.055$&$0.154\pm0.031$&$2.00^{+0.38}_{-0.65}$
&$1.285^{+0.031}_{-0.021}$&$1.347_{-0.093}^{+0.105}$
&$7.09\pm0.49$&$8.69_{-0.60}^{+0.68}$\\
KOI-261&$5708\pm13$&$4.329\pm0.030$&$0.048\pm 0.019$&$6.52^{+0.42}_{-0.35}$
&$1.057^{+0.009}_{-0.010}$&$1.165_{-0.046}^{+0.045}$
&$0.62_{-0.62}^{+1.09}$&$3.83_{-0.17}^{+0.16}$\\
KOI-262&$6150\pm53$&$3.994\pm 0.100$&$-0.104\pm 0.047$&$3.27^{+0.60}_{-0.52}$
&$1.374^{+0.091}_{-0.083}$&$1.963_{-0.257}^{+0.280}$
&$10.58\pm0.22$&$12.16_{-2.19}^{+2.86}$\\
KOI-269&$6371\pm50$&$4.160\pm 0.090$&$0.011\pm 0.049$&$2.21^{+0.21}_{-0.30}$ 
&$1.382^{+0.080}_{-0.062}$&$1.616_{-0.193}^{+0.227}$
&$11.62\pm0.22$&$15.28_{-1.85}^{+2.19}$\\
KOI-279$^\star$&$6531\pm58$&$4.425\pm 0.100$&$0.313\pm 0.063$&$<0.17$ 
&$1.450^{+0.022}_{-0.028}$&$1.258_{-0.022}^{+0.110}$
&$12.14\pm0.27$&$3.08_{-0.17}^{+0.27}$\\
KOI-280&$6047\pm40$&$4.262\pm 0.080$&$-0.258\pm 0.034$&$5.16^{+0.69}_{-0.90}$ 
&$1.095^{+0.034}_{-0.024}$&$1.279_{-0.123}^{+0.141}$
&$3.52\pm0.50$&$4.12_{-0.62}^{+0.79}$\\
KOI-302&$6616\pm63$&$3.882\pm 0.100$&$0.097\pm 0.054$&$1.33^{+0.15}_{-0.13}$ 
&$1.773^{+0.110}_{-0.117}$&$2.524_{-0.350}^{+0.398}$
&$16.66\pm0.16$&N/A\\
KOI-367&$5667\pm20$&$4.279\pm 0.040$&$0.151\pm 0.031$&$6.95^{+0.77}_{-0.79}$ 
&$1.071^{+0.016}_{-0.013}$&$1.243_{-0.059}^{+0.055}$
&$1.04\pm0.74$&$2.27_{-0.28}^{+0.36}$\\
KOI-974&$6385\pm30$&$4.058\pm 0.050$&$-0.026 \pm 0.031$&$2.27^{+0.12}_{-0.14}$ 
&$1.464^{+0.051}_{-0.047}$&$1.871_{-0.133}^{+0.145}$
&$7.13\pm0.49$&$8.74_{-0.63}^{+0.69}$\\
KOI-1463&$6578\pm70$&$3.886\pm 0.105$&$0.018\pm 0.069$&$1.48^{+0.12}_{-0.17}$ 
&$1.725^{+0.122}_{-0.116}$&$2.479_{-0.359}^{+0.416}$
&$9.66\pm0.36$&$20.76_{-3.00}^{+3.49}$\\
\hline
\end{tabular}
\end{table*}

Once $T_\mathrm{eff}$, $\log g$, and [Fe/H] are given, we can estimate
the stellar age, mass $M_s$, and radius $R_s$ for each system. We here
employ the Yonsei-Yale (Y$^2$) isochrone model to estimate these
parameters \citep{2001ApJS..136..417Y}.  The result is summarized in
Table \ref{table1}.  A visual inspection of the spectrum indicates that
KOI-1020 is a spectroscopic binary, which makes it difficult to derive
spectroscopic parameters.  In addition, the rotational velocities of
KOI-4, 100, 258, and 366, are so large ($V\sin I_s>30$ km s$^{-1}$) that
their spectra look very flat, which prohibits any reliable estimates of
the atmospheric parameters ($T_\mathrm{eff}$, $\log g$, $\xi$, and
[Fe/H]) with our current SNR's.  The best-fit values of $V\sin I_s$ for
those systems are estimated as 39.0 km s$^{-1}$, 32.8 km s$^{-1}$, 134.5
km s$^{-1}$, and 34.1 km s$^{-1}$, for KOI-4, 100, 258, and 366,
respectively. For such rapid rotators, asymmetries in the transit lightcurve 
may be used to determine the parameters only if the spin-orbit
angle is large \citep{2009ApJ...705..683B}. Such an analysis is beyond
the scope of this paper, and we do not perform it here.

\begin{figure}[t]
\begin{center}
\includegraphics[width=8.5cm,clip]{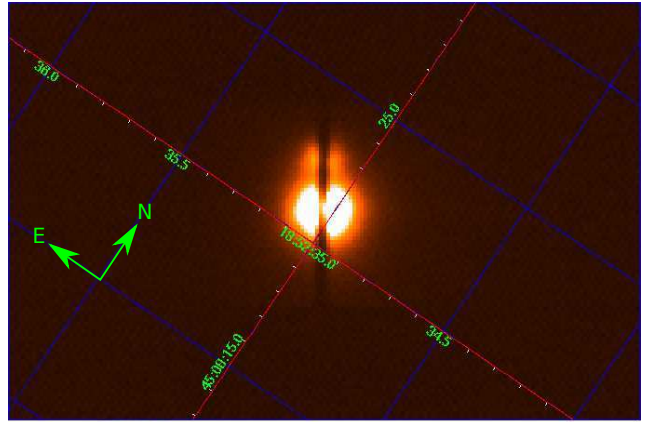} 
\caption{The slit viewer image for KOI-42. A companion star is seen about $2.0^{\prime\prime}$ 
to the north-east of the main star. The companion star is on the slit. 
}\label{fig:koi42}
\end{center}
\end{figure}
\begin{figure}[t]
\begin{center}
\includegraphics[width=8.5cm,clip]{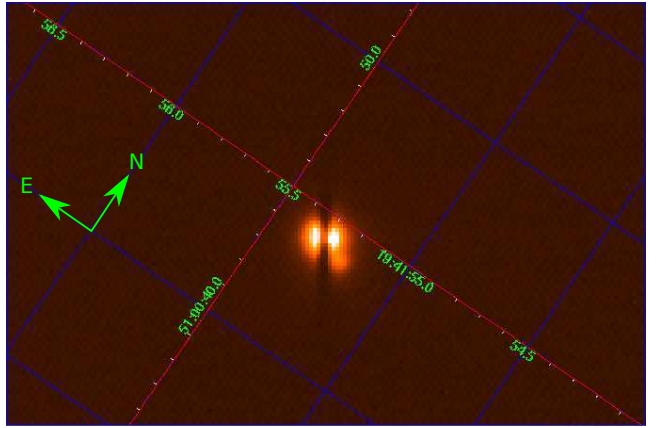} 
\caption{The slit viewer image for KOI-279. A companion star is seen about $1.5^{\prime\prime}$ 
to the west of the main star. The light from the companion is partly on the slit. 
}\label{fig:koi279}
\end{center}
\end{figure}

Regarding KOI-42 and KOI-279, the estimated values of $\log g$ seem too
large for stars with $T_\mathrm{eff}$ of $\sim6500$ K \citep[e.g., see Figure
2 in][]{1998A&A...338..161F}. These large values of $\log g$ may
explain the unusually young estimates for system ages listed in Table
\ref{table1}.  This tends to be caused by blending of light by a
companion star or another background source. Thus, we checked the slit
viewer images for those targets taken simultaneously with the spectra,
and consequently we found stellar companions for both of the systems and
those companions seem to be on the slit of Subaru/HDS, causing
significant contaminations in the spectra of KOI-42 and KOI-279 (see
Figure \ref{fig:koi42} and \ref{fig:koi279}) Kepler's photometric
aperture is relatively large, which means that these companion stars are
a big source of contamination.  This increases the chance that the
rotational periods estimated by spots are biased and might reflect the
companions' rotational periods.  Therefore, we simply ignore these two
systems in the following analysis and discuss the stellar inclinations
for the rest of the systems.

We also checked the slit viewer images for the other targets to see if
any contamination sources are located around the main objects. 
Consequently, we found that KOI-258 seems to have a companion star, 
located $\sim 1^{\prime\prime}$ to the east of the main object. 
This is consistent with our expectation
from the photometric analysis in Section \ref{s:periods_estimate}.
In addition, the point spread function (PSF) of 
KOI-1020 looked distorted, which suggests existence of a companion
or background sources. We could not locate any companion 
(within $\sim 1^{\prime\prime}$) nor anomalous PSF for the other systems. 

We compare our result for the stellar parameters in Table \ref{table1}
with the public KIC parameters, which are based on the photometric analyses.
We find that the root-mean-square (RMS) differences between our result and KIC values
for $T_\mathrm{eff}$ and $\log g$ are 217 K and 0.231 dex, respectively. 
These values are reasonably in good agreement with the reported uncertainties
for the KIC parameters
\citep[$\sim 200$ K and $\sim 0.4$ dex, respectively,][]{2011AJ....142..112B}. 
Moreover, when we remove the possibly contaminated systems 
(KOI-42, 279,  and 1463), the RMS differences significantly improve and
become 101 K and 0.134 dex for $T_\mathrm{eff}$ and $\log g$, respectively.

\subsection{Evidence of Possible Spin-Orbit Misalignments}

\begin{figure}[t]
\begin{center}
\includegraphics[width=9cm,clip]{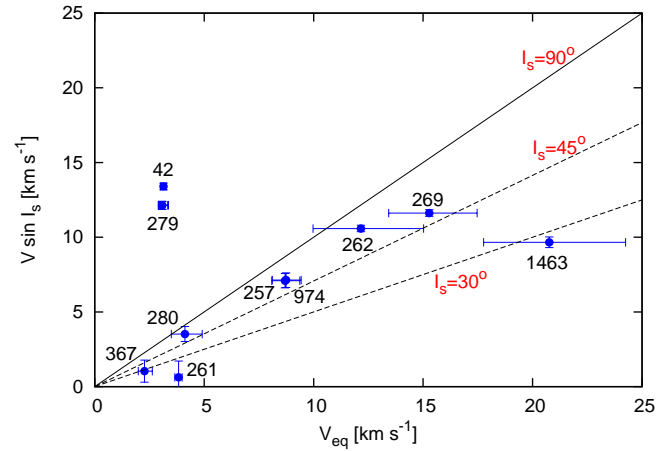} 
\caption{
The estimated $V_\mathrm{eq}$ and $V\sin I_s$. The solid line indicates
the case that our line-of-sight is vertical to the stellar spin axis. 
We also plot the two inclined cases in which $I_s=45^\circ$ and $30^\circ$.
}\label{fig:v_vsini}
\end{center}
\end{figure}

After excluding the systems for which we could not estimate either the
rotational period or the stellar radius, we are left with KOI-42, 257,
261, 262, 269, 279, 280, 367, 974, 1463.  KOI-42 and KOI-279 were
excluded as well because of the contaminations by companion stars, as
mentioned above.  In the right column of Table \ref{table1}, we also
show the rotational velocity at the stellar equator $V_\mathrm{eq}$
based on $P_s$ obtained by the Kepler photometry and spectroscopically
measured $R_s$.  In Figure \ref{fig:v_vsini}, we plot the projected
rotational velocity $V\sin I_s$ as a function of $V_\mathrm{eq}$.  In
this plot, we also show the three lines indicating $I_s=90^\circ$,
$I_s=45^\circ$, and $I_s=30^\circ$. While KOI-262 and KOI-280 are
consistent with $I_s=90^\circ$, the other systems are inconsistent with
$I_s=90^\circ$ within $1\sigma$. In particular, KOI-261 and KOI-1463
have significantly small stellar inclinations.  We note that KOI-262 is
a multiple transiting system (candidate), which makes the system a very
important sample to discuss planetary migrations \citep{2012Natur.487..449}.

As for the KOI-1463 system, for which we inferred a small $\sin I_s$,
the effective temperature of the host star ($\gtrsim 6500$ K) seems too
high to have dark starspots, as mentioned in Section
\ref{principle}. This means that the periodic flux variation of KOI-1463
may be spurious, 
implying that the flux variations do not originate
from the main star (with $T_\mathrm{eff}>6500$ K),
although it is still possible that smaller
spots (or spots with a higher brightness) induced the flux variation.  
Indeed the planet-to-star size ratio of KOI-1463 is reported to be $R_p/R_s=0.13655$ from the
Kepler transit lightcurve, and the radius of the transiting companion
becomes $36.93_{-6.21}^{+5.34} R_\oplus$ when we simply substitute the stellar radius
estimated via spectroscopy.  An object with such a huge radius 
corresponds to a very late-type star rather than a planet.
If this is the case, the periodic flux variation may come from the late-type companion
star, KOI-1463.01, which may well be active enough to have starspots.
Although the huge discrepancy between the public KIC parameter, which reports 
$R_s=1.17R_\odot$, and our estimate further supports this scenario, 
further lightcurve
analyses and/or a high resolution imaging are required in order to understand the
origin of the flux variation of KOI-1463.

While $I_s$ is just an angle between the stellar spin axis and the
line-of-sight, a small value of $I_s$ (or $\sin I_s$) for transiting
systems 
implies 
a spin-orbit misalignment, but not vice versa.
To see this more clearly, let us consider the 3D spin-orbit angle $\psi$
using Equation (\ref{eq:psi}) again. Recall first that the orbital
inclination $I_o$ is expressed as
\begin{eqnarray}
\cos I_o = b\frac{R_s}{a_p}\left(\frac{1+e\sin \varpi}{1-e^2}\right),
\label{eq:Io}
\end{eqnarray}
where $b$, $a_p$, $e$, $\varpi$ are the transit impact parameter,
planet's semi-major axis, orbital eccentricity, and longitude of the
periastron.  Except for extremely eccentric planets ($e\gtrsim 0.9$), we
expect that the right-hand-side of Equation (\ref{eq:Io}), apart from the
factor $b$, is of the order of $R_s/a_p$ and $b$ varies from zero to
unity by definition. Therefore,
\begin{eqnarray}
\cos I_o\lesssim \frac{R_s}{a_p}.
\label{eq:Ioconstraint}
\end{eqnarray}
Substituting the above relation into Equation (\ref{eq:psi}), we obtain
\begin{eqnarray}
\cos\psi&=&\sin I_s\cos\la\sin I_o+\cos I_s\cos I_o\nonumber\\
&\lesssim& \sin I_s+\frac{R_s}{a_p}\cos I_s.
\label{eq:psi_constrain}
\end{eqnarray}
Equation (\ref{eq:psi_constrain}) gives a lower limit of $\psi$ from
the observed value of $I_s$.  For instance, in the case of KOI-261, we
obtain $\psi\gtrsim 61^\circ$ based on Equation
(\ref{eq:psi_constrain}).

On the other hand, $I_s=90^\circ$ (i.e., $\sin I_s=1$) within its
uncertainty does not necessarily mean a spin-orbit alignment. This is
analogous to the case of $\la=0^\circ$ for an RM measurement, which does
not always imply a spin-orbit alignment. Statistical treatments are
important in both cases in order to compare observed distributions with
planetary migration theories.

\section{Discussion \label{s:discussion}}\label{s:discussion}
\subsection{Correlation between Stellar Inclinations and Other System Parameters}
\begin{figure}[t]
\begin{center}
\includegraphics[width=9cm,clip]{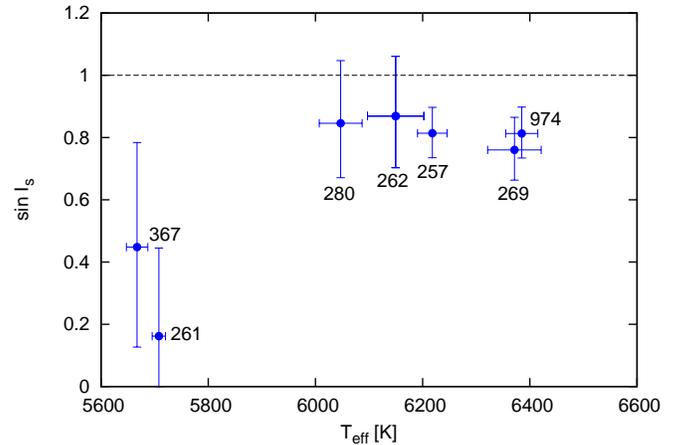} \caption{Correlation
between the stellar inclination and stellar effective temperature.  
}\label{fig:sini1-6}
\end{center}
\end{figure}
\begin{figure}[t]
\begin{center}
\includegraphics[width=9cm,clip]{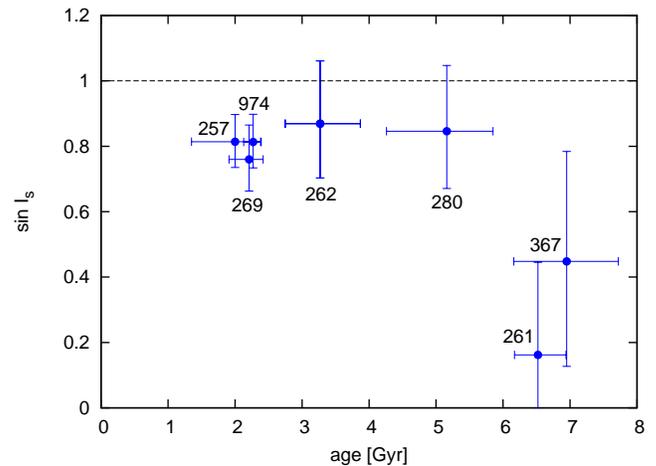} 
\caption{
Correlations between the stellar inclination and 
stellar age. 
}\label{fig:sini1-7}
\end{center}
\end{figure}

In the previous section, we have shown that at least one system (KOI-261) out of our 15
targets indeed indicates a possible spin-orbit misalignment along the line-of-sight. 
KOI-1463 also indicates a small inclination, but given the facts that its companion
is more likely a star rather than a planet and its effective temperature is too
high for KOI-1463 to have starspots, the inferred rotational velocity at the stellar equator
$V_\mathrm{eq}$ is suspicious. Therefore, we do not further consider the KOI-1463 
system with currently available data. 
Other systems (KOI-257, 269, 974)
also show possible spin-orbit misalignments with $\gtrsim 2\sigma$, 
but those moderate spin-orbit misalignments may be caused by stellar differential 
rotations as we will show later in Section \ref{sec:diff_rot}. 
Before discussing the impact of differential rotation, we here discuss the dependences 
of $I_s$ (that tells us about spin-orbit relations) on the system parameters. 

Figure \ref{fig:sini1-6} and \ref{fig:sini1-7} plot the stellar inclinations $\sin I_s$ against 
the stellar effective temperatures $T_\mathrm{eff}$ and their ages, respectively.  
As mentioned in Section \ref{s:intro}, host stars' effective temperatures and
ages are reported to have significant correlations with stellar obliquities 
\citep{2010ApJ...718L.145W, 2011A&A...534L...6T}. 
Amongst all, 
KOI-261 does not seem to follow the possible patterns by 
\citet{2010ApJ...718L.145W} and \citet{2011A&A...534L...6T}, 
who suggested that spin-orbit misalignments are seen around hot, young stars,
although KOI-261 has only a Neptune-sized planetary candidate
as shown later in this subsection. 



\begin{table*}[t]
\caption{
Correlation between $\sin I_s$ and Planetary Parameters.
}\label{table2}
\begin{center}
\begin{tabular}{cccccc}
\hline
Planetary Candidate &$\sin I_s$ & $P_o$ (days) (adopted) & $R_p/R_s$ (adopted) 
& $a_p$ (AU) & $R_p$ ($R_\oplus$) \\\hline\hline
KOI-257.01&$0.814_{-0.079}^{+0.083}$&$6.883403\pm0.000012$&$0.02052\pm0.00015$
&$0.0769^{+0.0007}_{-0.0004}$&$3.01^{-0.24}_{+0.21}$\\
KOI-261.01&$0.162_{-0.162}^{+0.283}$&$16.238480\pm0.000019$&$0.02431\pm0.00033$
&$0.1278\pm0.0004$&$3.09^{-0.12}_{+0.13}$\\
KOI-262.01&$0.869_{-0.166}^{+0.192}$&$7.8125124\pm0.000052$&$0.01074\pm0.00015$
&$0.0856_{-0.0017}^{+0.0019}$&$2.29^{-0.33}_{+0.30}$\\
KOI-262.02&$0.869_{-0.166}^{+0.192}$&$9.376137\pm0.000056$&$0.01362\pm0.00030$
&$0.0967^{+0.0021}_{-0.0020}$&$2.91^{-0.42}_{+0.38}$\\
KOI-269.01&$0.760_{-0.097}^{+0.105}$&$18.01134\pm0.00022$&$0.01074\pm0.00019$
&$0.1497_{-0.0022}^{+0.0029}$&$1.89^{-0.27}_{+0.23}$\\
KOI-280.01&$0.846_{-0.175}^{+0.201}$&$11.872914\pm0.000023$&$0.01972\pm0.00073$
&$0.1050_{-0.0008}^{+0.0010}$&$2.75^{-0.32}_{+0.28}$\\
KOI-367.01&$0.448_{-0.321}^{+0.336}$&$31.578680\pm0.000018$&$0.0420\pm0.0038$
&$0.2000_{-0.0008}^{+0.0010}$&$5.69^{-0.59}_{+0.57}$\\
KOI-974.01&$0.813_{-0.079}^{+0.085}$&$53.50607\pm0.00061$&$0.01353\pm0.00014$
&$0.3155_{-0.0034}^{+0.0037}$&$2.76^{-0.21}_{+0.20}$\\
KOI-1463.01&$0.465_{-0.069}^{+0.080}$&N/A&$0.13655$
&N/A&$36.93^{-6.21}_{+5.34}$\\
\hline
\end{tabular}
\end{center}
\end{table*}

\begin{figure}[t]
\begin{center}
\includegraphics[width=9cm,clip]{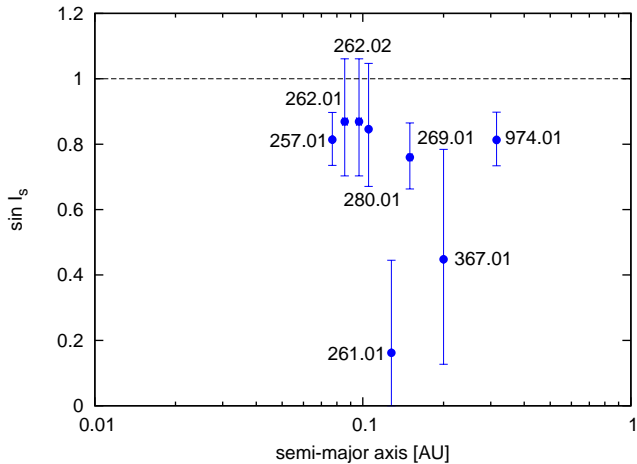}
\caption{Correlation between the stellar inclination and 
semi-major axes of planets.  
}\label{fig:sini2-8}
\end{center}
\end{figure}
\begin{figure}[t]
\begin{center}
\includegraphics[width=9cm,clip]{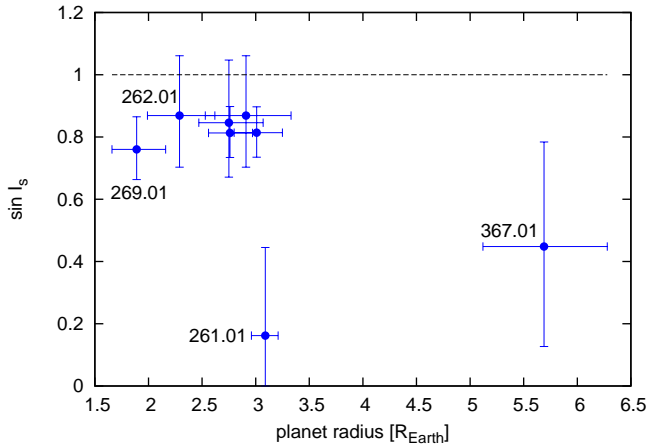} 
\caption{Correlation between the stellar inclination 
and planetary radii.  
}\label{fig:sini2-9}
\end{center}
\end{figure}
Next, in order to get a clearer insight into each of the
systems, we would like to provide a rough estimate for planetary parameters
by using the public photometric data. 
First, we focus on the semi-major axes of the planets. The distance between 
the planet and its host star plays a critical role in discussing planetary migrations, 
tidal interactions, and the resultant stellar obliquities with respect to planetary orbits. 
As usual, the planet semi-major axis $a_p$ is written as
\begin{eqnarray}
a_p=\{G(M_s+M_p)\}^{1/3}\left(\frac{P_o}{2\pi}\right)^{2/3},
\label{eq:Ioconstraint}
\end{eqnarray}
where $G$ is the gravitational constant, and $P_o$ and $M_p$ are the
orbital period and planet mass, respectively.  Now that we have a better
estimate for the stellar mass $M_s$ from spectroscopy, we substitute
$M_s$ in Table \ref{table1} into the above equation.  Adopting the
public orbital period for each of the KOI planetary candidates from the
MAST archive (as of March 2012), and neglecting the term of $M_p/M_s$
($\lesssim 0.001$), we obtain the semi-major axis $a_p$ shown in Table
\ref{table2}.  We plot the stellar inclination $\sin I_s$ for each
planetary candidate as a function of the semi-major axis $a_p$ in Figure
\ref{fig:sini2-8}.  

Finally, we deal with the planet radius. From the transit lightcurve, it
is generally possible to estimate the size of the transiting planet
relative to the stellar radius, but since the systems on which we are
focusing are only candidates, we need to keep in mind that lightcurve
may be blended with the flux from a binary companion or other background
sources.  In particular, because of the relatively large aperture size
of the Kepler photometry, some of the systems may well be blended. 
To quantify the rejection level of blending scenarios, 
we estimated the upper limits
of the secondary peaks in the spectral lines. Assuming the same line profile, we 
tried to fit a secondary peak for
each of the observed spectra with a differing flux ratio 
and Doppler shift. As a result, we typically obtained rejection levels ($1\sigma$ upper limit) 
of 0.01 to 0.11 for the flux ratio between the two peaks. 
The rejection levels are worse for rapidly rotating stars (i.e., KOI-262, 269, 974, and 1463)
because of the shallow line profiles. 
Note that our spectra typically have an S/N of $\sim 100$, so that the 
determinations of the upper limit 
are dominated by the photon noise.
Further observations are required in order to rule out the binary scenario.
Here, we simply adopt the planet-to-star size ratios $R_p/R_s$ reported by the
Kepler team, and infer the planet radii using the spectroscopically
estimated stellar radii.

The result is also shown in Table \ref{table2} and its correlation with
$\sin I_s$ is plotted in Figure \ref{fig:sini2-9}.  Most of the
candidates seem to be super-Earths or Neptune-sized planets. 
As for the two Neptune-sized planets orbiting relatively
cool stars, the orbit of KOI-261.01 seems to be inclined with respect to
the spin axis of the host star, as in the case of the super-Neptune
HAT-P-11b, for which a significant spin-orbit misalignment around a cool
star was found through the RM measurements \citep{2010ApJ...723L.223W,
2011PASJ...63S.531H} and a precise photometric analysis of the effect of
spots in the transit light curves \citep{2011ApJ...743...61S}.  As we
have noted in Section \ref{s:intro}, giant planets (including
Neptune-sized ones) discovered inside of the snow line ($\sim$ a few AU)
should have experienced planetary migrations.  The possible spin-orbit
misalignments in the KOI-261 system suggests that in some of the systems
with close-in Neptune-sized planets, planet-planet scatterings or other
dynamical processes such as the Kozai cycles may have played important
roles during their formations and evolutions.

\subsection{Comparison with Empirical Estimates for $P_s$}
\begin{table}[t]
\caption{
Rotational periods estimated by the empirical relation.
}\label{table3}
\begin{center}
\begin{tabular}{cccc}
\hline
System &$P_s$ (days)& $P_{s, \mathrm{model}}$ (days) & Flux Variability (\%)\\\hline\hline
KOI-257&$7.846\pm0.052$&$6.56_{-0.79}^{+0.59}$&0.1511\\
KOI-261&$15.38\pm0.30$&$25.98_{-0.74}^{+0.75}$&0.3357\\
KOI-262&$8.171\pm1.218$&$6.51_{-1.82}^{+2.80}$&0.0520\\
KOI-269&$5.351\pm0.136$&$5.06_{-0.91}^{+0.95}$&0.0213\\
KOI-280&$15.78\pm2.12$&$20.33_{-2.47}^{+2.61}$&0.0318\\
KOI-367&$27.65\pm3.56$&$25.80_{-2.08}^{+1.94}$&0.1161\\
KOI-974&$10.83\pm0.12$&$4.26_{-0.41}^{+0.56}$&0.0355\\
KOI-1463&$6.042\pm0.042$&$1.64_{-1.64}^{+1.36}$&0.0378\\
\hline
\end{tabular}
\end{center}
\end{table}

We have estimated the stellar rotational period $P_s$ based on the
periodic analysis of the Kepler photometry. While the current method is
valid as long as the periodic flux variation comes from the spot on
the stellar surface, the variation may be due to some other sources such
as flux variations of background stars.  In order to check the
reliability of $P_s$ that we estimated, we apply the gyrochronological
method as employed by \citet{2010ApJ...719..602S}.  We simply adopt the
same empirical relation for the rotational period as a function of the
stellar mass $M_s$ and age.  Substituting $M_s$ and ages listed in
Table \ref{table1} into the empirical formula by
\citet{2010ApJ...719..602S}, we obtain the modeled rotational periods
for the KOI systems in Table \ref{table2}, for which secure
spectroscopic parameters are obtained.  The derived periods $P_{s,
\mathrm{model}}$ are listed in Table \ref{table3} along with the
observed rotational periods $P_s$.  The uncertainty in $P_{s,
\mathrm{model}}$ comes solely from the uncertainties in $M_s$ and age
inferred from spectroscopy, and does not include any systematics in the
empirical relation.

The comparison between $P_s$ and $P_{s, \mathrm{model}}$ indicates that
for most of the systems the rotational periods derived from stellar
spots are in reasonably good agreement with the empirical ones, while
KOI-261, 974, 1463 show some discrepancies. This suggests that the
estimations of either of $P_s$ or spectroscopic parameters may be wrong
for those systems. It should also be emphasized, however, that the
empirical model by \citet{2010ApJ...719..602S} is rather simplified, and
it is difficult to evaluate the systematic errors caused by adopting
that empirical relation.  For instance, we note that KOI-1463's host
star is so massive that the empirical model may have significant
systematics because of the lack of the sampled stars in that region \citep[see
Figure 2 in][]{2010ApJ...719..602S}.  

There is a non-zero probability that the flux variability is caused by a
background star, a probability that is higher, the lower the observed
variability is.  To investigate this, we estimate the flux variability
by taking the detrended mean normalized flux, and eliminating the 5\%
highest and the 5\% lowest values.  This step helps to remove the effect
of outliers or artifacts that might still remain after the detrending
process. The flux variability is then defined as the range of values of the flux.
In Table \ref{table3}, we show the values of the variability for each system. As
one can see, KOI-974 and KOI-1463 show a very low level of variability,
with a higher false positive probability, whereas KOI-261 is the most
active star, which shows that the period of rotation has been calculated
more robustly.
Note that assuming that the flux variation in KOI-1463 comes from the low-mass
companion star and that the flux ratio between them is 50 to 100, a 0.0368\%
variability in the total flux corresponds to 1.89\% to 3.78\% variability in the companion's flux, 
which seems to be reasonably caused by starspots on an active late-type star.

\begin{figure}[t]
\begin{center}
\includegraphics[width=9cm,clip]{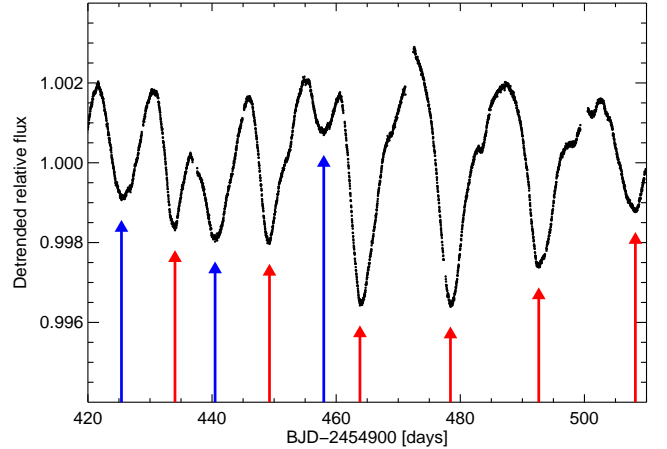} 
\caption{
The black dots represent the corrected flux series of KOI-261 on a segment of 90 days. 
In this segment, the star has a peak-to-peak variability of up to 0.6\%. The blue and 
red arrows point to the flux minima generated by two different active regions. 
The active region represented by blue arrows seems to disappear after three rotation 
periods, whereas the one represented by red arrows seems to reach its maximum size 
in the middle of the observations. In both cases, the flux minima recur with a periodicity 
of 15 days, confirming the value obtained with the Lomb-Scargle periodogram.
}\label{koi261_zoomed}
\end{center}
\end{figure}

Since KOI-261 is apparently the most important system in our sample 
that most likely to have a spin-orbit misalignment, we pay a special attention
to this system. 
If we adopt the modeled rotational period for KOI-261 as shown in Table \ref{table3},
we obtain $V_\mathrm{eq}=2.27\pm0.10$ km s$^{-1}$, which agrees with its projected 
rotational velocity of $V\sin I_s=0.62_{-0.62}^{+1.09}$ km s$^{-1}$ within $2\sigma$.
However, as we have shown in Figure \ref{koi261}, 
the periodogram clearly shows the single strongest peak around $15.5$ days.
The absence of a peak around $31$ days implies that the rotational period
is securely derived and does not reflect half the period. 
We also show in Figure \ref{koi261_zoomed} that there exist two active regions 
on the stellar surface manifested in KOI-261's lightcurve. 
The reason for the disagreement between the observed and modeled rotational periods
for KOI-261 is unknown, but it is general believed that such active stars as KOI-261 
rotate faster. Future confirmation and characterizations of this system is particularly
intriguing. 

\subsection{Impact of Differential Rotation\label{sec:diff_rot}}
So far, we have discussed the stellar inclinations $I_s$ assuming no
differential rotation of the planet hosting star. In reality, however, stars
may have differential rotations, which are supposed to be related to the origin of starspots, 
and the rotational velocity at the equator may be different from that estimated at the location of the star
spot.  Here, we discuss the impact of stellar differential rotations on
our estimate of $I_s$.

Following \citet{2003A&A...398..647R}, we model the angular velocity
$\Omega$ as a function of the latitude $l$ on the stellar surface as
\begin{eqnarray}
\Omega(l)=\Omega_\mathrm{eq}(1-\alpha\sin^2 l),
\label{eq:diffrot}
\end{eqnarray}
where $\Omega_\mathrm{eq}$ is the angular velocity of the star at the
equator.  The degree of differential rotation $\alpha$ is about 0.2 for
the case of the Sun.  In the presence of a differential rotation, it is
more explicit to rewrite Equation (\ref{I_s}) as
\begin{eqnarray}
\sin I_s=\frac{(V\sin I_s)_\mathrm{spec}}{R_s\Omega_\mathrm{eq}(1-\alpha\sin^2 l)}.
\label{I_s_diff}
\end{eqnarray}
Since we do not have information on $l$ where the stellar spot that we
observed is located, the uncertainty for $l$ results in a systematic
error for $\sin I_s$. A fortunate situation is, however, the
spectroscopically measured $V\sin I_s$ in the numerator of Equation
(\ref{I_s_diff}) is likely estimated to be smaller in the presence of
differential rotations ($\alpha>0$). This is because the solar-type
differential rotation leads to a ``sharper" spectral line profile
\citep{2003A&A...398..647R}, which results in a smaller value for $V\sin
I_s$ when fitted by the rotational broadening kernel for ``rigid
rotation".  Since the denominator of Equation (\ref{I_s_diff}) also
becomes smaller in the presence of differential rotation, the impact of
differential rotation tends to be more or less compensated.

In order to quantitatively evaluate Equation (\ref{I_s_diff}), we
perform the following simple numerical simulation. First, we generate
the mock line profile by convolving a single Gaussian function with a
rotational broadening kernel (including macroturbulence) for a
differentially rotating star. The input parameters to create one line
profile are the rotational velocity at the equator $V_\mathrm{eq}\equiv
R_s\Omega_\mathrm{eq}$, differential rotation parameter $\alpha$, and
input stellar inclination $(\sin I_s)_\mathrm{in}$\footnote{ The other
parameters such as the original Gaussian width and macroturbulence
dispersion are fixed in the simulation.}.  Then, we fit the resultant
line profile with a convolution function between a Gaussian and a
rotational broadening kernel for a rigid body, and estimate the best-fit
value for $(V\sin I_s)_\mathrm{spec}$.

\begin{figure}[t]
\begin{center}
\includegraphics[width=9cm,clip]{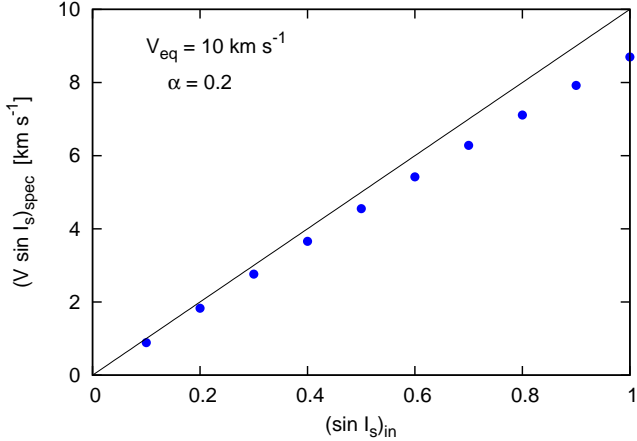} 
\caption{The simulated result for the line profile fitting in the
presence of differential rotation. We plot the best-fit values for
$(V\sin I_s)_\mathrm{spec}$ obtained by fitting the line profiles of
differential rotation with the rotational kernel for a ``rigid
rotation'', for various input stellar inclinations $(\sin
I_s)_\mathrm{in}$.  In this plot, we employ $V_\mathrm{eq}=10$ km
s$^{-1}$ and $\alpha=0.2$, and the black line represents the expected
$V\sin I_s$ for rigid rotation.  The best-fit results for $(V\sin
I_s)_\mathrm{spec}$ are deviated from the cases for rigid rotation by
$\sim 10\%$.  }\label{fig:diff_fit}
\end{center}
\end{figure}

As a result, in the case of $\alpha=0.2$, the best-fit values for
$(V\sin I_s)_\mathrm{spec}$ are smaller than the input values of
$V_\mathrm{eq}\cdot(\sin I_s)_\mathrm{in}$ by $\sim 10\%$, for various
cases of $(\sin I_s)_\mathrm{in}$ (see Figure \ref{fig:diff_fit}).  This
result implies that changing $\sin l$ from 0 to 1 in Equation
(\ref{I_s_diff}) leads to the systematic errors in the output $\sin I_s$
of $\sim \pm 10\%$.
We check it for various values of the input parameters $V_\mathrm{eq}$,
$\alpha$, and $(\sin I_s)_\mathrm{in}$ and find that the relative
systematic error in estimating $\sin I_s$ from Equation (\ref{I_s_diff})
is approximately $0.5\alpha$.

In the case of the Sun, however, the latitudes at which we observe the
spots are confined to the relatively narrow bands of $5^\circ\lesssim
|l|\lesssim 40^\circ$. The active latitude is known to move toward the
solar equator with a cycle of $\sim 11$ years, making the well-known
``butterfly diagram'' \citep[e.g.,][]{2005SoPh..229...35R}.  

\begin{figure}[t]
\begin{center}
\includegraphics[width=9cm,clip]{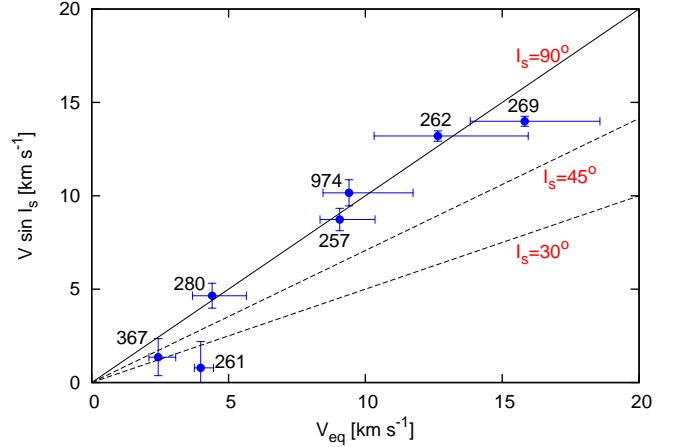} 
\caption{
Same as Figure \ref{fig:v_vsini}, after corrected for the impact
of differential rotations. 
}\label{fig:v_vsini_diff}
\end{center}
\end{figure}
In order to correct for the impact of differential rotation, we here adopt the
following empirical model for $\alpha$ reported by \citet{2007AN....328.1030C}:
\begin{eqnarray}
\alpha \Omega_\mathrm{eq}=0.053(T_\mathrm{eff}/5130)^{8.6}.
\label{eq:diffrot}
\end{eqnarray}
This expression, which is based on observations of
Doppler imaging or Fourier analyses of the rotational broadening kernel, claims that 
differential rotations are stronger for hotter stars such as our targets
\citep{2007AN....328.1030C, 2008MNRAS.387.1525D}.
We perform the same simulation using the mock line profile
in the presence of differential rotation for each of the seven systems, 
and then correct for the impact of differential rotation. 
We evaluate the systematic
errors in $V_\mathrm{eq}$ assuming that the spots are located at 
$|l|=20^\circ\pm 20^\circ$, imitating the case of the Sun. 

Figure \ref{fig:v_vsini_diff} plots thus corrected 
$V_\mathrm{eq}$ - $V\sin I_s$ diagram. 
In contrast to Figure \ref{fig:v_vsini}, KOI-257, 269, 367, and 974 are consistent
with $I_s=90^\circ$ with $\sim 1\sigma$, while KOI-261 still seems to have
a spin-orbit misalignment. This implies that differential rotations can 
partially or totally explain the moderate and apparent spin-orbit
misalignments for KOI-257, 269, 367, and 974, although we do not conclude so 
in the present paper because of the rather rough assumptions
we adopted for the correction of differential rotation. 
We note that if this is the case, it is equally interesting in the sense that the present method 
can reveal the signature of differential rotations of distant transiting planetary systems.

\subsection{
Comparison with the RM Measurement for Kepler-8\label{sec:kepler8}
}
Finally, we briefly discuss the application of the present technique to the confirmed
Kepler systems. 
Specifically, systems where the RM effect has been measured are particularly
interesting since we can directly compare between the sky-projected 
and line-of-sight spin-orbit angles.
Among the confirmed Kepler systems with the RM measurements, 
Kepler-8 shows a periodic flux variation, and we obtain the rotational period of
$P_s=7.5\pm0.3$ days. This period, along with the stellar radius of $R_s=1.486_{-0.062}^{+0.067}R_\odot$
\citep{2010ApJ...724.1108J}, results in $V_\mathrm{eq}=10.0\pm0.6$ km s$^{-1}$, which 
is in good agreement with the projected rotational velocity $V\sin I_s=10.5\pm 0.7$ km s$^{-1}$
\citep{2010ApJ...724.1108J}. 
Although $I_s\approx 90^\circ$ is only a requirement for a spin-orbit alignment, 
this result does not support the RM measurement by \citet{2010ApJ...724.1108J}
($\lambda=-26.4^\circ\pm 10.1^\circ$). 
See also \citet{2012IAUS..282..379A}, who suspected 
the uncertainty of $\lambda$ for Kepler-8 to be underestimated. 

\section{Summary and Future Prospects \label{s:conclusion}}\label{s:conclusion}

We have investigated the stellar inclinations for KOI planetary system
candidates on the basis of the detailed analyses of photometric
variation due to stellar spots and spectroscopic measurements of $V\sin
I_s$. We have found that at least one system, KOI-261, exhibits a strong
signature of a possible spin-orbit misalignment along the line-of-sight.
The planetary candidate KOI-261.01 is a Neptune-sized one ($R_p\approx 3 R_\oplus$), 
with a moderate orbital distance ($a_p\approx0.13$ AU). 
If this system is confirmed and eventually turns out to be misaligned, KOI-261.01 
will become the smallest planet ever reported to have a spin-orbit misalignment, 
which makes it an important sample to test and discuss planetary migrations in 
an extended parameter space. 
KOI-1463 also shows a small stellar inclination, though 
the size of the transiting companion may correspond to that of an M star
and the periodic signal may come from the companion. 
The results for the other systems are ambiguous but they may be interpreted 
as either mildly misaligned or differentially rotating.

One of the next tasks is to increase the number of samples and further
discuss the correlations between $I_s$ and other system parameters, for
each of the Earth-sized planet population and giant planet population.
This will require to choose fainter stars, unless new updates of the
catalog bring many candidates around variable hot stars.  Even with the
currently available dataset, 
it is relatively easy to plan a more efficient observing run, giving higher
priority to stars with a higher variability and ruling out those ones
with no variability or with signals that are too coherent to be caused
by spots. 

It is also important to refine the measurements of $I_s$.  We are
potentially able to do so if we combine the spectroscopic measurement
with photometric data.  For instance, the ratio of the planet semi-major
axis to stellar radius is available via transit photometry, but we did
not use it in estimating the stellar radius $R_s$. This is simply
because $a/R_s$ generally depends on the orbital eccentricity $e$, and
$e$ cannot be determined by the transit alone unless the secondary
eclipse is seen in the lightcurve.  If radial velocities for the systems
presented in this paper are followed-up by some future observations and
their orbital eccentricities are measured, we are able to constrain
$R_s$ more precisely and the estimates for $\sin I_s$ would be
significantly improved.  

Finally, we stress that all the systems that we analyzed are still candidates for 
having planetary companions, and therefore we must be careful in
discussing the evolution history of planetary systems with the present technique.
However, we can safely say that possible misaligned systems suggested by our analysis 
would become very interesting targets for future RM measurements. 

\acknowledgments 

This paper is based on data collected at Subaru Telescope, which is
operated by the National Astronomical Observatory of Japan.  
We are very grateful to Simon Albrecht, for helpful comments
on this manuscript. 
We acknowledge the support for our Subaru HDS observations by Akito
Tajitsu, a support scientist for the Subaru HDS.  T.H.\ 
expresses special thanks to Akihiko Fukui for a fruitful discussion on
this subject.  The data analysis was in part carried out on common use
data analysis computer system at the Astronomy Data Center, ADC, of the
National Astronomical Observatory of Japan.  T.H.\ is supported by Japan
Society for Promotion of Science (JSPS) Fellowship for Research (DC1:
22-5935).  J.N.W.\ acknowledges support from the NASA Origins program
(NNX11AG85G).  
T.H. and N.N. acknowledge a support by NINS Program for
Cross-Disciplinary Study.
N.N. is supported by Grant-in-Aid for Research Activity Start-up No. 23840046.
Y.S. gratefully acknowledges support from the
Global Collaborative Research Fund ``A World-wide Investigation of Other
Worlds'' grant and the Global Scholars Program of Princeton University,
the Grant-in-Aid No. 20340041 by JSPS, and JSPS Core-to-Core Program
``International Research Network for Dark Energy''.  We wish to
acknowledge the very significant cultural role and reverence that the
summit of Mauna Kea has always had within the indigenous people in
Hawai'i. We 
express special thanks to the anonymous referee for the 
helpful comments and suggestions on this manuscript.





\end{document}